# Robust valley-polarized excitonic Mott states and doublons enabled by stacking-controlled moiré geometry

*Hao-Tien Chu*[†1], *Shou-Chien Chiu*[†1], *Meng-Che Yeh*[†1], *Yu-Wei Hsieh*[†1], *Jia-Sian Su*[1], *Xiao-Wei Zhang*[2], *Jie-Yong Zeng*[1], *Po-Chun Huang*[1], *Si-Jie Chang*[1], *Kenji Watanabe*[3], *Takashi Taniguchi*[4], *Yunbo Ou*[5], *Seth Ariel Tongay*[5], *Ting Cao*[*2], *Chaw-Keong Yong*[*1,6]

[1] Department of Physics, National Taiwan University, Taipei 10617, Taiwan

[2] Department of Materials Science and Engineering, University of Washington, Seattle, WA, USA.

[3] Research Center for Electronic and Optical Materials, National Institute for Materials Science, 1-1 Namiki, Tsukuba 305-0044, Japan

[4] Research Center for Materials Nanoarchitectonics, National Institute for Materials Science, 1-1 Namiki, Tsukuba 305-0044, Japan

[5] Materials and Engineering, School for Engineering of Matter, Transport and Energy, Arizona State University, Tempe, Arizona 85287, United States.

[6] Center of Atomic Initiative for New Materials, National Taiwan University, Taipei 10617, Taiwan

† These authors contributed equally to this work.

* Correspondence to: tingcao@uw.edu, chawkyong@phys.ntu.edu.tw

**Abstract**

Atomically-thin moiré superlattices offer an optically accessible platform for interacting bosons, where strong onsite repulsion $U_{xx}$ suppresses double occupancy and supports excitonic Mott states at unit filling. However, moiré confinement also enhances phonon- and disorder-assisted relaxation, challenging the robustness of these correlated states under dissipation. Here we show that strengthening the intersite exciton repulsion $V_{xx}$ between neighboring moiré cells offers a distinct route to stabilizing unit-filling excitonic Mott states. In H-stacked $WSe_2/WS_2$, moiré confinement endows interlayer excitons with an out-of-plane dipole and a pronounced in-plane quadrupolar charge distribution. Helicity-resolved transient photoluminescence, supported by first-principles–informed modelling, reveals that this quadrupolar geometry increases $V_{xx}$ at unit filling by at least a factor of two relative to the dipolar R-stacked excitons. Despite a slight reduction in $U_{xx}$, the enhanced $V_{xx}$ yields a long-lived, valley-polarized excitonic Mott state at unit filling that persists for ~12 ns—more than twice as long as in R-stacks—and remains robust up to ~50 K. Beyond unit filling, the same geometry supports valley-polarized doublons with fourfold longer lifetimes than in R-stacks. These results establish moiré-geometric control of intersite interactions as a route to stabilizing excitonic Mott states and doublons against dissipation in solids.

Two-dimensional (2D) semiconducting moiré heterobilayers, such as $WSe_2/WS_2$ and $MoSe_2/WSe_2$, provide a powerful platform for tailoring correlated excitonic phases at the nanoscale. Type-II band alignment yields long-lived interlayer excitons[1–4], while twist-induced moiré superlattices generate flat minibands that suppress tunneling and enhance Coulomb interactions[5–11]. These ingredients have enabled a range of collective phenomena, including excitonic Mott incompressibility[1,2,5,6,12], Wigner crystal states[7,10,13], stripe order[14], and light-induced magnetism[15], establishing moiré semiconductors as solid-state analogues of interacting quantum lattices.

Recent studies further demonstrate that stacking registry can reshape moiré exciton wavefunctions[6,16,17]. Fig. 1a summarizes the first-principles results for $WSe_2/WS_2$: R-stacked (0° twist) heterobilayers host compact dipolar interlayer excitons with vertically aligned charge centers, whereas H-stacked (60° twist) heterobilayers laterally displace the electron and hole, producing a spatially extended exciton with an out-of-plane dipole and a pronounced in-plane quadrupolar component. The exciton-exciton interactions in moiré exciton lattices are often framed within an extended Bose-Hubbard framework (Fig. 1b), where strong onsite repulsion $U_{xx}$ suppresses double occupancy near unit filling ($\bar{v}_{ex} = 1$) and intersite repulsion $V_{xx}$ penalizes departure from the unit-filling configuration in an excitonic Mott state[1,5,6]. Because the spatial extent and charge distribution of the exciton wavefunction set both local overlap and Coulomb coupling, stacking-controlled moiré geometry can in principle reshape both onsite and intersite interactions (Fig. 1c), and thereby control the formation and robustness of correlated many-body exciton dynamics. Establishing this connection is challenging because exciton–phonon coupling and disorder-assisted scattering rapidly drive relaxation and can obscure the underlying interaction fingerprints on nanosecond timescales[3,18,19].

Here, we combine helicity-resolved transient photoluminescence (PL) with first-principles-informed modelling to show that stacking-controlled moiré geometry stabilizes correlated exciton-lattice dynamics by enhancing intersite repulsion. In H-stacked $WSe_2/WS_2$, the in-plane quadrupolar character of the interlayer exciton increases the effective $V_{xx}$ at unit filling by more than a factor of two, while slightly reducing the onsite double-occupancy cost $U_{xx}$ relative to the dipolar excitons in R-stacks. Strikingly, the enhanced $V_{xx}$ yields a long-lived, valley-polarized excitonic Mott state at unit filling that persists for ~ 12 ns – more than twice as long as in R-stacks – and remains robust up to ~50 K. Beyond unit filling, the same geometry supports valley-polarized doublons with more than fourfold longer lifetimes, delaying relaxation into the unit-filling state. These results establish stacking-controlled intersite interactions as practical route to tailor nonequilibrium correlated excitons in solids: $U_{xx}$ sets the unit-filling Mott constraint, whereas enhanced $V_{xx}$ controls the robustness of unit-filling Mott states against dissipation and prolongs valley-polarized doublon dynamics.

**Moiré excitons in WSe$_2$/WS$_2$ heterobilayers**

We fabricated high-quality WSe$_2$/WS$_2$ moiré superlattice heterostructures by mechanical exfoliation and deterministic stacking following Ref. [20]. Monolayer flakes were assembled with controlled twist angles of either 0° (R-stack) or 60° (H-stack), encapsulated by hexagonal boron nitride (hBN) layers, and transferred onto an alumina-coated silver mirror (Methods). The twist angle between WSe$_2$ and WS$_2$ layers was verified by polarization-resolved second-harmonic generation (Supplementary Section 1). Differential reflectance spectroscopy at 4 K reveals three sharp absorption peaks near the WSe$_2$ A-exciton resonance (Fig. 1d), which serve as the spectral fingerprint of spatially uniform moiré-confined exciton states[8,17,21]. Figure 1b summarizes the relevant energy scales of the moiré exciton lattice, including the onsite doublon cost $U_{xx}$, intersite repulsion $V_{xx}$, tunneling $t$, and exchange splitting $\Delta_{ex}$. As we show below, stacking registry reshapes these Hubbard parameters – most prominently the intersite repulsion $V_{xx}$ and the effective onsite overlap – leading to distinct nonequilibrium Mott and doublon dynamics under dissipation.

To resolve the correlation-driven processes before dissipation dominates, we excite the samples using circularly polarized femtosecond pulses ($\sigma^+_{\text{pump}}$) resonant with the WSe$_2$ A-exciton and probe the interlayer exciton emission at a delay time $\Delta\tau = 1$ ns using helicity-resolved transient PL spectroscopy (~0.7 ns temporal resolution; Methods). This delay is early enough that intersite tunneling and nonradiative loss remain negligible, yet late enough for excitons to relax fully into moiré potential minima[4,22,23], consistent with the transient reflectance lineshape becoming time-independent beyond ~100ps (Extended Data Fig. 1). Figure 1e shows helicity-resolved PL spectra for both moiré geometries at $\Delta\tau$ = 1 ns, each exhibiting pronounced exciton emission with strong valley-polarization. Furthermore, the valley contrast becomes diminishingly weak in time-integrated PL (Extended Data Fig. 2), highlighting the essential role of time-resolved spectroscopy in revealing the valley and correlation dynamics of moiré excitons.

## Density-dependent photoluminescence at $\Delta\tau$ = 1 ns

At a fixed delay $\Delta\tau = 1$ ns, we map how the interaction landscape evolves with filling. Below $\bar{\nu}_{\text{ex}} = 1$, $V_{\text{xx}}$ produces a gradual blueshift of emission with increasing density (Supplementary Section 2). Once above $\bar{\nu}_{\text{ex}} = 1$, onsite repulsion produces an abrupt energy jump of magnitude $U_{\text{xx}}$ [1,5,6]. Helicity-resolved detection further separates interaction-driven energy shifts from valley-dependent exchange effects. Fig. 2a,b show the helicity-resolved emission from H-stacked and R-stacked heterobilayers at $\Delta\tau = 1$ ns, respectively. Below unit filling ($\bar{\nu}_{\text{ex}} < 1$; excitation power < 70 nW, Methods), both devices exhibit a single emission line, IX$_1$, arising from singly-occupied moiré sites. Increasing density induces a continuous blueshift of IX$_1$, reflecting repulsive intersite excitonic interactions $V_{\text{xx}}$. Near $\bar{\nu}_{\text{ex}} = 1$, the IX$_1$ energy shift ($\Delta E_{(\bar{\nu}_{\text{ex}}=1)}$) relative to the dilute limit (extrapolated to zero pump power) reaches ~ 5 meV in R-stacks but ~ 11 meV in H-stacks (Fig. 2g,h), indicating substantially stronger intersite Coulomb interactions in the H-stacked exciton lattices.

Once $\bar{\nu}_{\text{ex}} > 1$, IX$_1$ saturates and a second emission line, IX$_2$, emerges – approximately 36 meV above IX$_1$ in R-stacks and 27 meV in H-stacks (Supplementary Section 3). This discrete splitting reflects the onsite interaction energy $U_{\text{xx}}$ required to form a doublon – a spectroscopic hallmark of the bosonic Mott state in WSe$_2$/WS$_2$ moiré heterostructures[1,5,6]. The IX$_2$ intensity increases much more slowly with excitation density in R-stacks than in H-stacks (Fig. 2c,d), suggesting rapid annihilation of dipolar doublons and highlighting fundamentally distinct many-body dynamics dictated by moiré geometry. Tracking the peak energies with increasing density reveals how adding excitons above unit filling reshapes the correlation landscape. In R-stacks, both IX$_1$ and IX$_2$ quickly plateau in energy, indicating that interactions become effectively onsite once each moiré cell is occupied (Fig. 2h). In contrast, H-stacks exhibit a continued blueshift of both branches deep into $\bar{\nu}_{\text{ex}} > 1$, demonstrating that intersite repulsion continues to evolve in the over-filled lattice despite strong onsite repulsion $U_{\text{xx}}$ (Fig. 2g).

Helicity-resolved measurements further sharpen this contrast. While IX$_1$ remains strongly valley-polarized in both systems, IX$_2$ retains robust valley polarization in H-stacks but shows no measurable valley contrast in R-stacks (Fig. 2e,f). The vanishing valley contrast in R-stacks is consistent with a large onsite exchange splitting ($\Delta_{\text{ex}} \sim 30$ meV) expected for dipolar doublons, which favors opposite-valley pairing[5]. By contrast,

IX$_2$ in H-stacks shows clear valley contrast, implying suppressed exchange-driven depolarization within the doublon manifold. A ~2 meV helicity splitting of IX$_2$ (Fig. 2g) further indicates a substantially reduced effective exchange splitting. The reproducibility of this behavior across multiple device pairs (Extended Data Fig. 3) confirms that stacking registry governs both the doublon interaction pathways and the valley configuration of excitons on doubly-occupied sites.

At longer delays ($\Delta\tau = 6$ ns), exciton annihilation and tunneling increasingly reshape the many-body landscapes (Extended Data Fig. 4). While IX$_1$ remains largely unchanged, IX$_2$ vanishes entirely in R-stacks but remains observable in H-stacks, albeit with reduced intensity and valley polarization. The IX$_2$-IX$_1$ energy separation in H-stacks also decreases to ~ 23 meV, indicating a dynamic softening of $U_{xx}$ as tunneling and diffusion begin to compete with onsite repulsion.

Taken together, the density-dependent blueshifts and doublon energetics show that moiré geometry strongly modifies intersite interactions. Figure 2i summarizes $U_{xx}$ and the intersite-induced blueshift of IX$_1$ at unit filling, $\Delta E_{(\bar{\nu}_{ex}=1)}$. Although $U_{xx}$ reduces slightly in H-stacks, $\Delta E_{(\bar{\nu}_{ex}=1)}$ is nearly twice as large as in R-stacks, signifying a strong enhancement of intersite repulsion $V_{xx}$. As we show below, this stacking-controlled interaction landscape arises from the moiré-controlled exciton wavefunction and ultimately governs the correlated dynamics under dissipation.

**Microscopic origin of enhanced intersite repulsion**

To uncover the microscopic origin of stacking-dependent many-body excitonic interactions, we computed the moiré electronic structure using density functional theory (DFT), including full structural relaxation for each moiré geometry (Methods). From the relaxed structures, we extracted the spatial wavefunction of the moiré valence-band maximum and conduction-band minimum, corresponding to the frontier electron and hole orbitals that host the interlayer excitons (Supplementary Section 4). The resulting moiré flat bands exhibit pronounced spatial modulation that depends sensitively on the moiré geometry. In R-stacked heterobilayers, both electrons and holes localize at the same $R_h^X$ site, yielding interlayer excitons in which the electron in the WS$_2$ layer is vertically aligned with the hole in the WSe$_2$ layer (Supplementary Section Fig. S4). The exciton ground state therefore carries a predominantly out-of-plane electric dipole. In contrast, H-stacked heterobilayers display a qualitatively different charge distribution: the hole remains localized at the $H_h^X$ site in WSe$_2$, whereas the electron wavefunction spreads over three neighboring $H_h^h$ sites in WS$_2$ with threefold rotational symmetry, yielding an interlayer exciton with both an out-of-plane dipole and a pronounced in-plane quadrupolar moment.

We first examine the regime above unit filling ($\bar{\nu}_{\text{ex}} > 1$) by evaluating the energetic cost of adding an exciton to the moiré superlattice already filled at $\bar{\nu}_{\text{ex}} = 1$. Screened Coulomb energies were computed using the stacking-dependent exciton charge distribution obtained from first-principles calculations (Supplementary Section 5). In R-stacked heterobilayers, adding a second exciton to the same $R_h^X$ site produces a purely onsite repulsion $U_{\text{xx}} = 36$ meV, accompanied by negligible coupling to neighboring moiré sites (Fig. 3a). This locally screened interacting picture is consistent with experiment, which shows a density-independent $U_{\text{xx}} = 36$ meV (Fig. 3f) and a vanishingly small blueshift of IX$_1$ ($\Delta E_{\bar{\nu}_{\text{ex}} > 1}$) above unit filling (Fig. 3e).

In H-stacks, by contrast, the in-plane quadrupolar charge distribution allows the added exciton to interact not only with its onsite partner but also with excitons in neighboring moiré cells. This quadrupole-enabled intercell interaction channel in the two-exciton manifold (Fig. 3b) is characterized by an onsite interaction energy $U_{\text{xx}}$ of 27 meV and an additional intersite contribution, $E_{\text{xx}}^{(2)}$, of 2.8 meV between the doubly occupied site and surrounding singly-occupied sites. Such quadrupole-mediated extended interactions parallel those in extended Bose-Hubbard settings realized in Rydberg-dressed atomic lattices, where interactions beyond onsite terms reshape the properties of repulsively bound pairs[24–26]. Crucially, here the extended interaction emerges intrinsically from the moiré-modified exciton wavefunction rather than from externally engineered dressing fields. As doublon density increases, more singly-occupied sites fall within the range of these quadrupole-mediated couplings (Fig. 3c,d). To compare directly with experiment, we spatially average the doublon-induced intersite interaction energy experienced by singly-occupied sites, $\langle E_{\text{xx}}^{(2)} \rangle$, over a large periodic moiré supercell comprising 1000 x 1000 unit cells at varying doublon densities (Supplementary Section 5). The calculated trends agree well with experiments: $U_{\text{xx}}$ remains density-independent at 27 meV (Fig. 3f), whereas $\langle E_{\text{xx}}^{(2)} \rangle$ closely tracks the measured blueshift of IX$_1$ with increasing exciton filling

(Fig. 3e), supporting a picture in which quadrupolar excitons open an additional intercell interaction channel in the two-exciton manifold.

Moreover, our calculations show that $U_{xx}$ decreases with increasing in-plane charge separation $\Delta r_{CM}$ within doubly-occupied sites, reflecting reduced wavefunction overlap between the two excitons. This trend mirrors the experimentally observed temporal softening of $U_{xx}$ at later delays (Fig. 3f), when intersite tunneling and diffusion of doublons become increasingly relevant[3].

Below unit filling, the blueshift of IX$_1$ directly probes intersite Coulomb repulsion among singly-occupied moiré sites. Figure 3g compares the measured IX$_1$ blueshift with the spatially averaged intersite interaction energy, calculated as $\langle E_{xx}^{(1)} \rangle = \frac{1}{N_a} \sum_{a \neq b} V_{xx}(a,b)$, where $a$ and $b$ label distinct moiré sites and $N_a$ is the number of occupied sites (Supplementary Section 5). As the filling approaches unity, H-stacked heterobilayers exhibit a substantially larger blueshift than R-stacked devices, in good agreement with the corresponding enhancement of $\langle E_{xx}^{(1)} \rangle$. This larger $\langle E_{xx}^{(1)} \rangle$ in H-stacks originates from the quadrupolar exciton wavefunction discussed above (Fig. 1c), which strengthens intersite Coulomb coupling relative to compact dipolar excitons in R-stacks. Consequently, the intersite interaction energy build up much more rapidly as the filling approaches unity, accounting for the strong blueshift of IX$_1$ observed experimentally. At $\bar{\nu}_{ex} = 1$, $\langle E_{xx}^{(1)} \rangle$ reaches 11.7 meV in H-stacks, far exceeding the 3.8 meV in R-stacks. This pronounced enhancement of intersite repulsion raises the energetic cost of departures from unit filling and establishes a markedly stronger intersite interaction landscape near the unit-filling regime in H-stacks

**Nonequilibrium exciton dynamics: doublon relaxation and unit-filling Mott plateau**

The larger $V_{xx}$ in H-stacks raises a central question: does enhanced intersite repulsion make the unit-filling Mott regime more robust under dissipative conditions, even though $U_{xx}$ is slightly reduced? To address this question, we perform helicity-resolved time-gated PL measurements over extended delays following femtosecond excitation. We first analyze the relaxation dynamics of IX$_1$ and IX$_2$, which arise from singly- and doubly-occupied moiré sites, respectively. Fig. 4a,b compare the time-dependent decay traces of the IX$_2$ and IX$_1$ emission features for H-stacked and R-stacked devices at an excitation power of 115 nW ($\bar{\nu}_{ex} \approx 1.5$), extracted from helicity-resolved transient PL spectra by two-Lorentzian fitting. IX$_1$ decays slowly over hundreds of nanoseconds and retains strong helicity contrast over extended delays in both devices. By contrast, IX$_2$ decays within a few nanoseconds, and its dynamics differ strongly between the two stackings. In R-stacks, the IX$_2$ decays with a time constant of 1.7 ns and shows no measurable valley polarization. In H- stacks, the IX$_2$ decay time increases to 7.9 ns and retains detectable valley polarization over a similar timescale. These trends are consistent with reduced local two-exciton overlap in H-stacks, which weakens short-range loss and exchange-driven depolarization in the doublon channel.

To visualize how the system evolves into the near-unit-filling regime, we plot the total PL transients ($\sigma^+ + \sigma^-$) for several initial fillings, obtained by directly summing the $\sigma^+$ and $\sigma^-$ spectra at each delay, in Fig. 4c,d. For initial fillings above unity, the PL first decays rapidly as the doublon channel relaxes, then settles into a long-lived plateau. We assign this plateau to a unit-filling excitonic Mott state for three reasons: it is reached only after the doublon channel has decayed and the emission becomes dominated by IX$_1$; it appears immediately when the initial excitation density is tuned to prepare $\bar{v}_{\text{ex}} = 1$; and during the plateau the optical response becomes pinned. In particular, the IX$_1$ emission becomes nearly independent of the excitation density, consistent with saturation at the moiré site density, while the peak position and linewidth remain essentially unchanged throughout the plateau (Extended Data Fig. 5). These observations support a pinned, incompressible configuration in which double occupancy is blocked by the large onsite interaction $U_{\text{xx}}$, while redistribution away from unit filling is energetically disfavored by $V_{\text{xx}}$. Consistent with this interpretation, exciton diffusion is strongly suppressed on nanosecond timescales in this regime, as shown previously[3]. Meanwhile, the emission retain a large helicity contrast (Extended Data Fig. 6), revealing a transient valley-polarized Mott state. At longer delays, phonons and disorder melt the plateau, and the emission resumes a gradual decay dominated by the intrinsically slow IX$_1$ channel over hundreds of nanoseconds.

Strikingly, the plateau persists for more than ~12 ns in H-stacks, more than twice as long as in R-stacks. Its longer duration in H-stacks is consistent with the larger unit-filling interaction-induced shift $\Delta E_{(\bar{v}_{\text{ex}}=1)}$, supporting the view that enhanced intersite repulsion $V_{\text{xx}}$ strengthens the robustness of the excitonic Mott state under dissipation. Temperature-dependent measurements reinforce this picture: the plateau vanishes by 30

K in R-stacks (Fig. 4f), but persists to higher temperatures in H-stacks and is fully suppressed only by 50 K (Fig. 4e). This stacking-dependent thermal robustness supports a central role for intersite interactions $V_{xx}$ in stabilizing excitonic Mott states under dissipative conditions. The driven-dissipative evolution at 4 K for H-stacked and R-stacked devices is summarized in Fig. 4g,h, highlighting how moiré geometry controls both the formation and robustness of the excitonic Mott plateau.

**Conclusion**

Correlated bosonic dynamics in solids are inherently nonequilibrium and compete with phonon-, disorder-, and recombination-driven relaxation. In moiré exciton lattices, onsite repulsion $U_{xx}$ enforces the unit-filling constraint, whereas our results show that enhanced intersite repulsion $V_{xx}$ provides a complementary lever for stabilizing this correlated state under dissipation. By reshaping interlayer excitons to acquire pronounced in-plane quadrupolar character, H-stacked $WSe_2/WS_2$ strengthens intersite repulsion $V_{xx}$ between neighboring moiré cells and thereby extends both the lifetime and thermal robustness of the unit-filling excitonic Mott state. At the same time, reduced onsite overlap weakens exchange-driven depolarization and Auger loss, stabilizing long-lived valley-polarized doublons. More broadly, moiré-geometric control of intersite interactions, complementary to tuning onsite physics, provides a route to extended-interaction bosonic regimes in moiré exciton lattices, where interaction range, dissipation, and valley degrees of freedom can be co-designed. Combined with electrical tuning[27], dielectric screening[19], and valley/topological band design[28,29], this approach may provide access to new quantum phases of matter and establish moiré exciton lattices as a broadly tunable platform for correlated bosons.

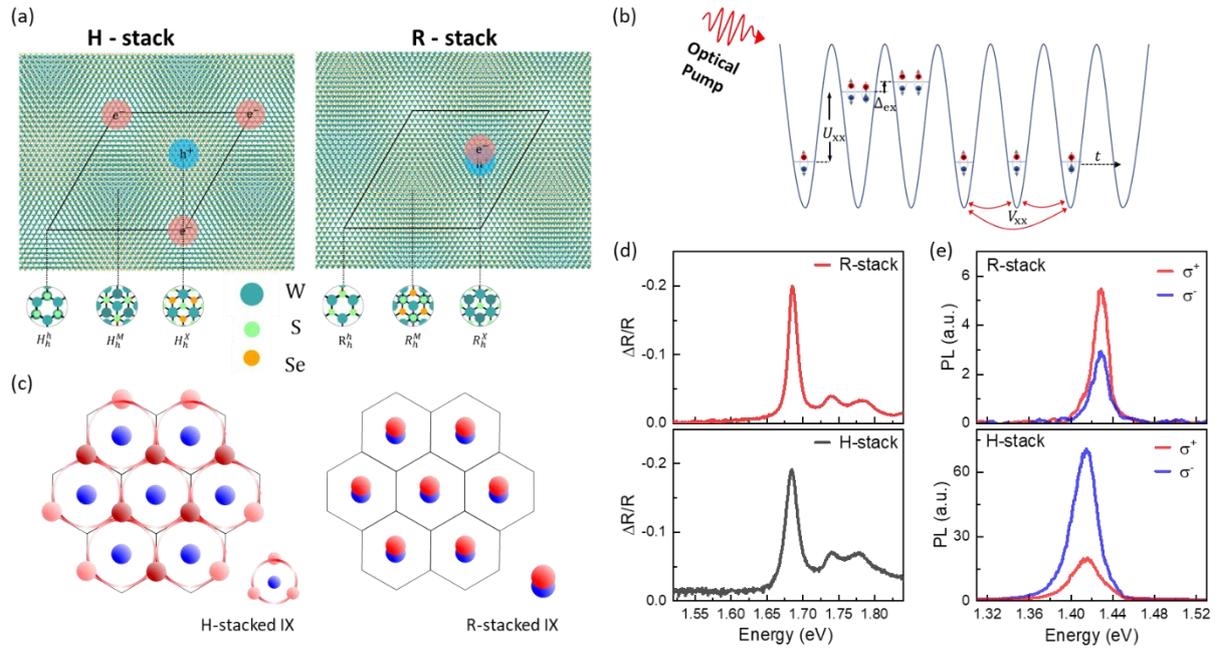

**Figure 1| Twist-tailored interlayer moiré excitons in WSe$_2$/WS$_2$ heterobilayers. a.** Illustration of the spatial distributions of band edge electron (red) and hole (blue) in H-stacked and R-stacked WSe$_2$/WS$_2$ heterobilayers. Insets show the atomic registry of the three high-symmetry sites within a moiré supercell. High symmetry labels are defined with respect to the hollow site (h) of the WSe$_2$ layer (subscript), while the superscript (h, M (W), or X (S)) denotes the WS$_2$ atomic site vertically aligned with the WSe$_2$ hollow center. In R-stacks, the electron in WS$_2$ is vertically aligned with hole in WSe$_2$ at the $R_h^X$ site, forming a compact interlayer exciton with a dominant out-of-plane dipole moment. In H-stacks, the hole remains localized at the $H_h^X$ site in WSe$_2$, while the electron wavefunction spreads over three neighboring $H_h^h$ sites in WS$_2$, producing an interlayer exciton with both an out-of-plane dipole and a pronounced in-plane quadrupolar charge distribution. **b.** Schematic of key energy scales governing many-body excitonic interactions in moiré exciton lattices. **c.** Exciton confinement in moiré superlattices at $\bar{v}_{\text{ex}} = 1$. In H-stacks, the electron probability density is distributed over three $H_h^h$-type regions within each moiré cell, enhancing Coulomb coupling between excitons in neighboring cells. **d.** Differential reflectance at 4 K near the WSe$_2$ A-exciton resonance, showing three sharp absorption features of moiré-confined intralayer exciton states. **e.** Helicity-resolved PL spectra of interlayer excitons measured at $\Delta\tau = 1$ ns after $\sigma^+$ excitation, demonstrating strong valley-contrasting emission.

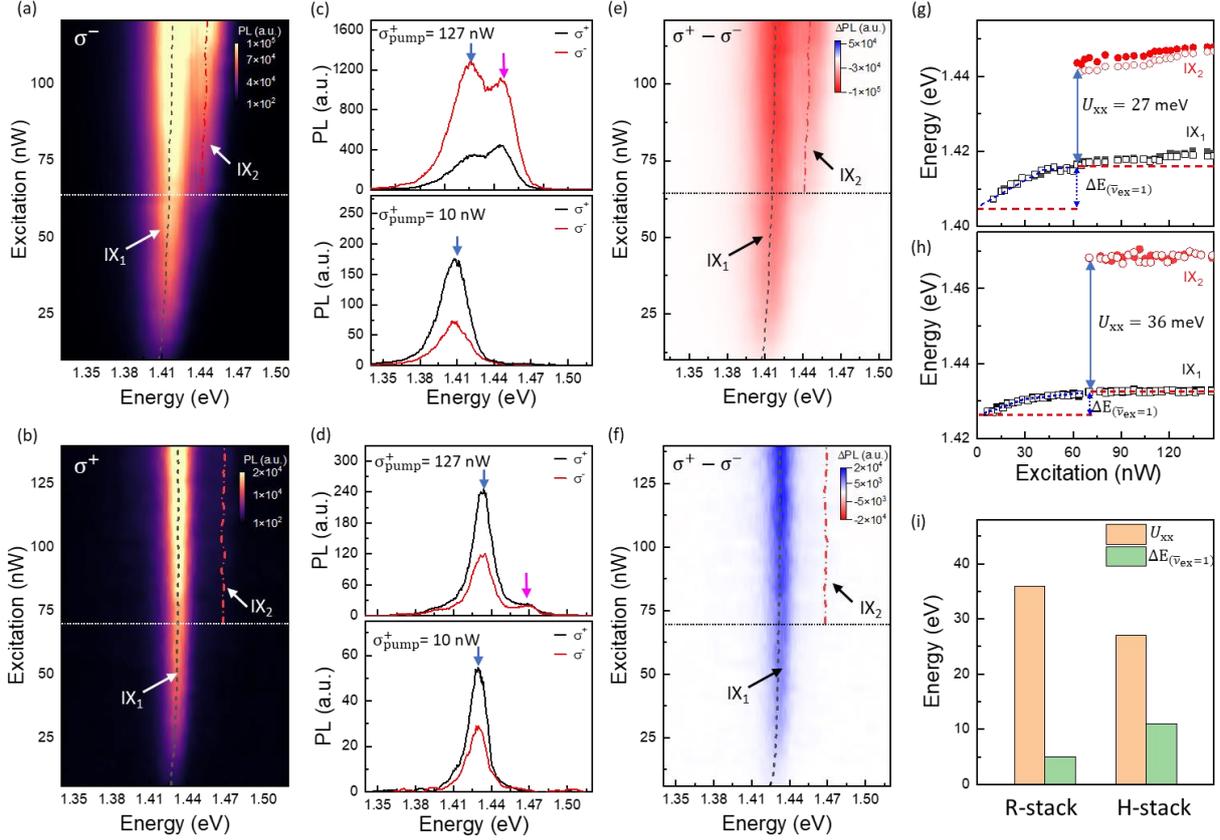

**Figure 2| Optical signatures of doublon formation and interaction-induced energy shift at $\Delta\tau = 1$ ns. a,b.** Helicity-resolved PL spectra for (a) H-stacked and (b) R-stacked WSe$_2$/WS$_2$, recorded at $\Delta\tau = 1$ ns. Increasing pump fluence drives a crossover from single-occupancy line (IX$_1$) to a higher-energy doublon line (IX$_2$). Horizontal dotted lines mark the $\bar{\nu}_{ex} = 1$ threshold; vertical markers track IX$_1$ and IX$_2$. **c,d.** Helicity-resolved PL spectra at selected excitation powers for (c) H-stacked and (d) R-stacked devices; blue and magenta arrows mark the peak positions of IX$_1$ and IX$_2$, respectively. **e,f.** Helicity contrast ($\sigma^+ - \sigma^-$) for (e) H-stacked and (f) R-stacked devices. IX$_1$ remains strongly polarized in both stackings, whereas IX$_2$ shows robust polarization only in H-stacks. **g,h.** Extracted peak energies versus excitation density for H-stacked (g) and R-stacked (h) devices. The onsite repulsion energy $U_{xx}$, obtained from the IX$_2 -$ IX$_1$ separation, is ~27 meV (H-stacks) and ~36 meV (R-stacks). $\Delta E_{(\bar{\nu}_{ex}=1)}$ denotes the blueshift of IX$_1$ at unit filling relative to the dilute limit (extrapolated to zero pump) and is ~11 meV (H-stacks) and ~5 meV (R-stacks). Open and solid symbols in (g,h) represent $\sigma^+$ and $\sigma^-$ detection. **i.** Summary of extracted $U_{xx}$ and $\Delta E_{(\bar{\nu}_{ex}=1)}$ for R-stacks and H-stacks.

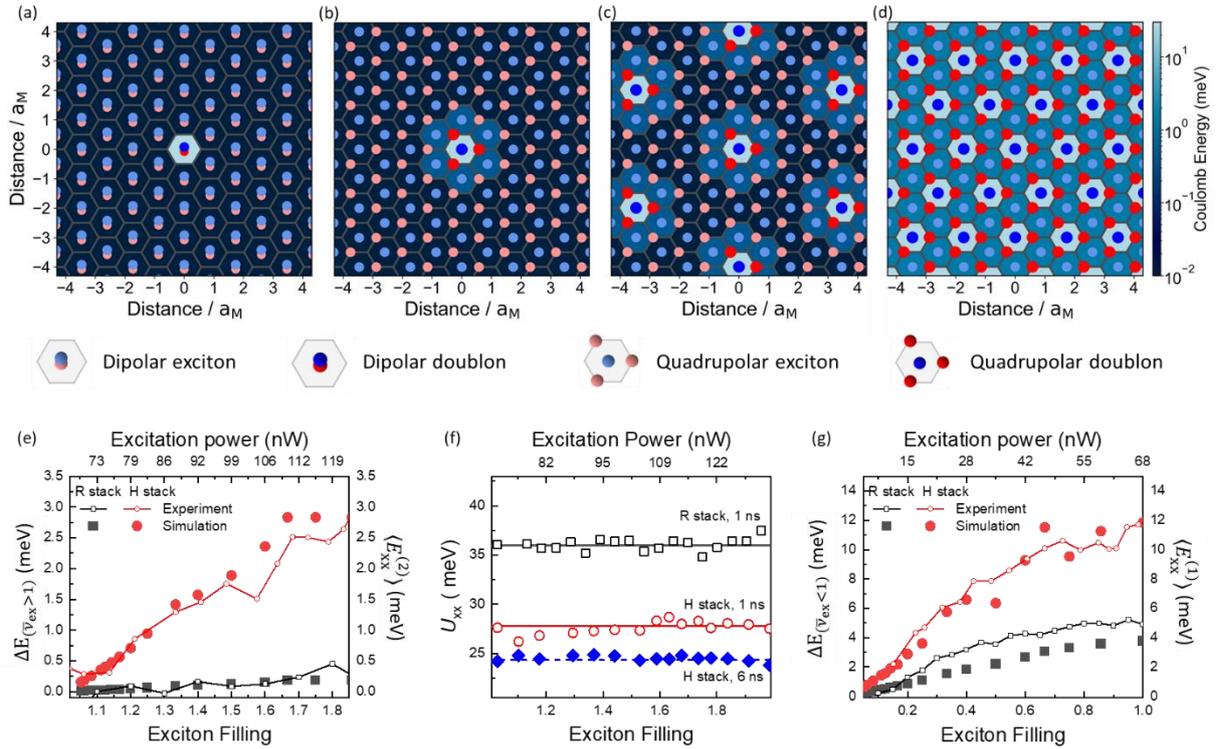

**Figure 3| Quadrupole-driven excitonic correlations. a-d.** Calculated Coulomb energy associated with adding one exciton to a moiré lattice already at unit filling in (a) R- and (b) H-stacked heterobilayers. In each moiré cell, the color scale denotes the Coulomb energy of an exciton whose center of mass is located in that cell. In R-stacks, compact dipolar excitons produce a purely onsite repulsion of ~36 meV with negligible coupling to neighboring cells. In H-stacks, the quadrupolar excitons wavefunction opens an additional intercell interaction channel in the two-exciton manifold, giving an onsite energy of ~27 meV together with an extra doublon–neighbor coupling of ~2.8 meV. **c,d.** Calculated local Coulomb-energy maps in H-stacks for increasing filling, from (c) $\bar{v}_{\text{ex}} = 17/16$ to (d) $\bar{v}_{\text{ex}} = 4/3$. As the doublon density increases, a larger fraction of singly-occupied sites falls within the range of quadrupole-mediated intercell couplings. **e.** IX$_1$ energy shift $\Delta E_{(\bar{v}_{\text{ex}}>1)}$ for $\bar{v}_{\text{ex}} > 1$, referenced to the emission energy at unit filling $\bar{v}_{\text{ex}} = 1$ (solid lines with open symbols). R-stacks show negligible $\Delta E_{(\bar{v}_{\text{ex}}>1)}$, whereas H-stacks exhibit a continuous increase that is quantitatively captured by the calculated spatially averaged doublon-induced intersite energy $\langle E_{\text{xx}}^{(2)} \rangle$ (solid symbols). Top axis indicates excitation power. **f.** $U_{\text{xx}}$ as a function of filling, measured at 1 ns (open symbols) and 6 ns (closed symbols), compared with theory that includes wavefunction overlap within the moiré cell (solid and dashed lines). g. IX1 blueshift for $\bar{v}_{\text{ex}} < 1$ ($\Delta E_{(\bar{v}_{\text{ex}}<1)}$) referenced to the dilute-excitation limit. As the filling approaches unity, H-stacks exhibit a substantially

larger blueshift than R-stacks, in agreement with the calculated spatially averaged intersite energy $\langle E_{\text{xx}}^{(1)} \rangle$, reflecting quadrupole-enhanced intercell repulsion.

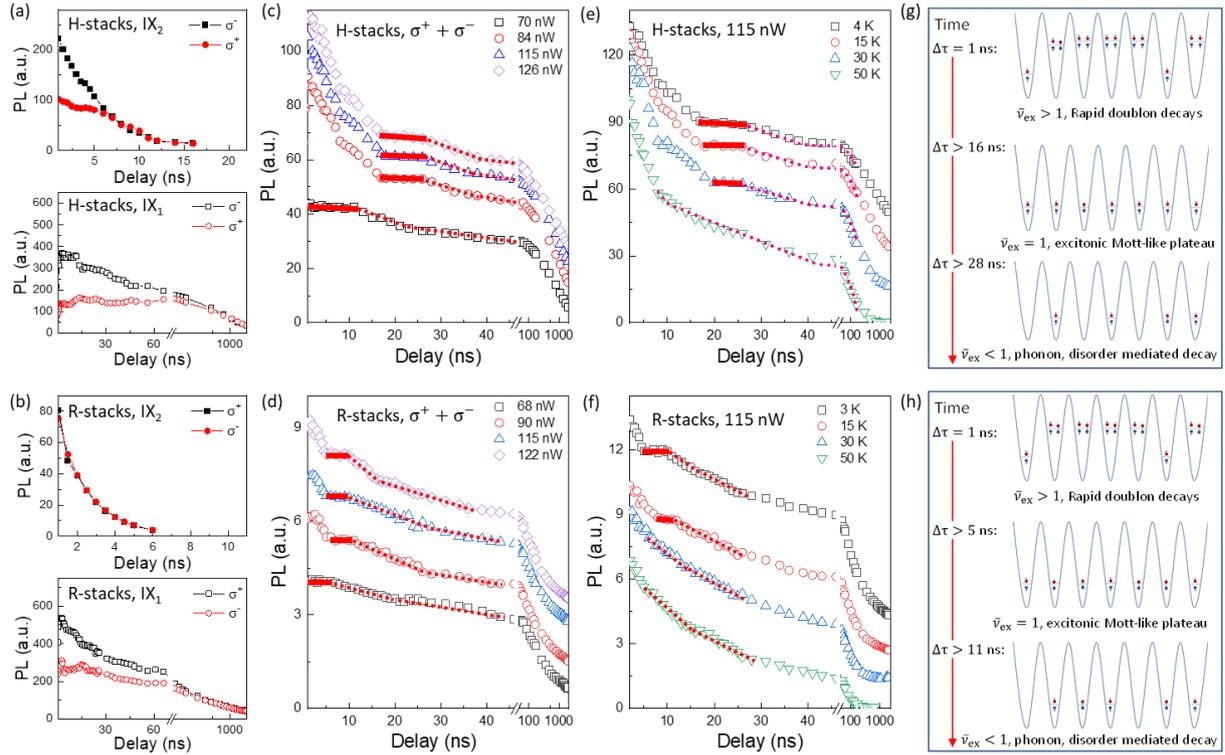

**Figure 4| Nonequilibrium doublon decay and emergence of the excitonic Mott plateau. a,b.** Time-dependent decay traces of the IX$_2$ and IX$_1$ emission features for (a) H-stacked and (b) R-stacked WSe$_2$/WS$_2$ heterobilayers at an excitation power of 115 nW ($\bar{\nu}_{ex} \approx 1.5$), extracted from helicity-resolved transient PL spectra by two-Lorentzian fitting. IX$_2$ decays much more slowly in H-stacks than in R-stacks. **c,d.** Total PL ($\sigma^+ + \sigma^-$) transient for (c) H-stacks and (d) R-stacks at several excitation densities, obtained directly by summing the $\sigma^+$ and $\sigma^-$ spectra at each time delay. Dotted lines guide the eye; horizontal bars mark the long-lived plateau dominated by IX$_1$ after rapid doublon decay. For initial filling above unity (excitation power > 70 nW), the plateau emerges after ~5 ns in R-stacks but is delayed to ~16 ns in H-stacks, reflecting the longer-lived doublon channel in the quadrupolar geometry. When the initial filling is tuned to $\bar{\nu}_{ex} = 1$, the plateau appears immediately. **e, f.** Temperature dependence of the total PL transient ($\sigma^+ + \sigma^-$) for (a) H-stack and (b) R-stack at 115 nW excitation. As in **c,d**, these traces are obtained directly from the summed helicity channels without spectral fitting. The plateau shortens with increasing temperature and vanishes near 50 K in H-stacks and 30 K in R-stacks. Traces are vertically offset for clarity; dotted lines guide the eye and horizontal bars indicate the plateau interval. **g,h.** Schematic of the driven-dissipative evolution in (g) H-stacked and (h) R-stacked heterobilayers following $\sigma^+$ excitation, illustrating rapid doublon decay above unit filling, formation of the unit-filling Mott plateau, and its eventual melting by phonon- and disorder-assisted relaxation.

## Methods

**Sample Fabrication.** The WSe$_2$/WS$_2$ heterobilayers encapsulated in hBN flakes were assembled using a dry-transfer technique with a polypropylene carbonate (PPC) stamp[20]. Monolayer WSe$_2$, WS$_2$, and hBN flakes were first exfoliated onto silicon substrate with a 285 nm thermal oxide. Crystal orientations were determined by polarization-resolved second harmonic generation (SHG). The heterostructures were assembled by sequentially picking up the top hBN, monolayer WSe$_2$, WS$_2$, and bottom hBN layers with a PPC stamp, using optical alignment to control the stacking angle. The complete stacks were released onto a silver substrate coated with an 85 nm Al$_2$O$_3$ layer. During transfer, the polymer stamp and substrate were heated to 50 °C for pick-up and 85 °C for release. After fabrication, polarization-resolved SHG was performed on both monolayer regions of the sample to determine the exact twist angle and on the heterostructure region to distinguish between near-zero and near-60° samples (Supplementary Section 1). Samples were annealed at 200 °C for 6 h in an Argon flow chamber.

**Time-resolved PL spectroscopy.** The heterostructures were mounted in a closed-cycle cryostat (Montana Instruments) with a temperature stability better than 10mK. Unless otherwise specified, all measurements were performed at 4 K. A custom-built optical microscope was used to record PL dynamics. Excitation pulses (∼ 120 fs) were generated by an optical parametric amplifier (OPA, ORPHEUS, Light Conversion) pumped by a 175 kHz regenerative amplifier (Light Conversion PHAROS). The beam was focused onto the sample and PL was collected in reflection geometry using a 20x objective (NA = 0.45, Olympus). Pump intensity was tuned with a variable neutral density filter (Thorlabs). Excitation and detection polarizations were independently controlled using a combination of polarizer, half-wave plate and quarter-wave plate. Transient PL spectra were recorded with an electronically-gated intensified charge-coupled device camera (iCCD, Andor iStar) coupled to a calibrated spectrometer (Andor SR500i). Nanosecond-scale temporal evolution was obtained by varying the gate delay of iCCD relative to the excitation pulse. The temporal resolution of the system is ∼ 700 ps.

**Differential Reflectance spectroscopy.** Differential reflectance measurements were carried out using the same optical setups as in time-resolved PL measurements. A supercontinuum probe (550-1100 nm) was generated by focusing a portion of the regenerative amplifier output onto a 5 mm thick sapphire crystal. The reflected signal was dispersed by a spectrograph and detected with a CCD camera (Andor iDUS 420).

**Transient Reflectance spectroscopy.** The pump-probe spectroscopy study is based on a regenerative amplifier seeded by a mode-locked oscillator (Light Conversion PHAROS). The regenerative amplifier delivers femtosecond pulses at a repetition rate of 175 kHz and a pulse duration of ∼ 150 fs, which were split into two beams. One beam was used to pump an optical parametric amplifier and the other beam was focused onto a sapphire crystal to generate supercontinuum light for probe pulses. The cross-correlation of the pump and probe pulses has a full-width half-maximum close to ∼200 fs. The pump-probe time delay was controlled by a motorized delay stage. The probe light was detected by a high-sensitivity CCD line camera operated at 75 Hz. The helicity of the pump and probe pulses were independently controlled using broadband quarter waveplates and linear polarizers. Both pump and probe pulses were focused onto the sample using a 20x objective (NA = 0.45, Olympus). Pump intensity was tuned with a variable neutral density filter (Thorlabs).

**Determination of exciton density.** Experimentally, the pump power was tuned using a variable neutral density filter, with excitation resonant to the $WSe_2$ A-exciton at 1.68 eV. The excitation spot radius was ∼ 2 $\mu$m. From Figure 2a,b, the crossover from single-occupancy line $IX_1$ to a higher-energy doublon line $IX_2$ occurs near a pump power of ∼70 nW, corresponding to an excitation fluence of 3.1 $\mu$J/cm². At this fluence, the incident photon density is 1.1 x 10¹³ cm⁻². The photon absorption was estimated from the differential reflectance spectrum in Fig. 1d: because the heterostructures were placed on an $Al_2O_3$ coated silver substrate, the reflectance loss directly reflects absorption[30]. At the A-exciton resonance, the absorption was ∼ 18%, giving an absorbed photon density of ∼ 2.1 x 10¹² cm⁻². Since charge transfer across the $WSe_2/WS_2$ heterostructure occurs within ∼ 50 fs, much faster than the radiative recombination of intralayer excitons, and the interlayer excitons thermalize into moiré traps within 100 ps, well before significant Auger recombination (Extended Data Fig. 1), we infer that the absorbed photon density provides a reliable estimate of the interlayer exciton density. The extracted density agrees closely with the expected value for unit filling $\bar{v}_{ex} = 1$, given by $n_o = \frac{2}{\sqrt{3}a_M^2} \approx 2.1 \times 10^{12}$ cm⁻², where $a_M \approx 7 - 8$ nm is the moiré period for the measured twist (device-specific values in Table S1).

**First-principles calculations of moiré structures and electronic structures.** The lattice relaxations of moiré superlattices are modeled using machine learning force fields (MLFFs) parameterized via the deep potential molecular dynamics (DPMD) method[31,32]. The training datasets are generated from 5500-step ab initio molecular dynamics (AIMD) simulations of 5.68° H-stacked and R-stacked $WSe_2/WS_2$ heterostructures, performed with the VASP package[33]. Van der Waals interactions are included through the D2 correction scheme[34]. The embedding and fitting neural networks each comprise three hidden layers, with a cutoff radius of 10.0 Å for each atom. The force root means square error (RMSE) is approximately 0.02 eV/Å, as evaluated on 100 independent 5.68° DFT configurations. The trained MLFFs are then used to relax the experimentally relevant aligned R- and H-stacked moiré structures in the LAMMPS package[35] until the maximum atomic force is below $10^{-4}$ eV/Å.

Electronic band structures of the relaxed moiré superlattices, including spin–orbit coupling (SOC), are calculated with the SIESTA package[36]. The calculations employ optimized norm-conserving Vanderbilt pseudopotentials, the Perdew–Burke–Ernzerhof (PBE) exchange–correlation functional[37], and a double-ζ plus polarization basis set.

## Acknowledgements

This work is primarily supported by the National Science and Technology Council (NSTC) in Taiwan under grant no. MOST 111-2636-M-002-024 (NSTC Young Scholar Fellowship Program) and Ministry of Education (MOE) in Taiwan under grant no. NTU-111V1011-2 (Yushan Young Scholar Fellowship Program). The research is also supported in part by the Higher Education Sprout Project by the Ministry of Education (MOE) in Taiwan under grant no. NTU-111L104047 and NTU-111L7886. S.T.A acknowledges direct support from DOE SC0020653 for excitonic metrology on monolayer 2D TMDs from vdW crystals, vdW crystals. S.A.T also acknowledges Applied Materials Inc. (materials synthesis), NSF CBET 2330110 (materials testingenvironmental testing). K.W. and T.T. acknowledge support from the JSPS KAKENHI (Grant Numbers 21H05233 and 23H02052), the CREST (JPMJCR24A5), JST and World Premier International Research Center Initiative (WPI), MEXT, Japan. The authors thank F. Wang for useful discussion. Hao-Tien Chu, Shou-Chien Chiu, Meng-Che Yeh, and Yu-Wei Hsieh, contributed equally to this work.

## Conflict of Interest

The authors declare no conflict of interest.

## Supplementary Information

Additional data including analysis of PL spectra, modelling of exciton-exciton interactions, and additional measurements (PDF).

## Author Contributions

The study was conceived by C.-K.Y. C.-K.Y., H.-T.C, J.-S.S., S.-C.C, M.-C.Y, and Y.-W.H designed the experiments, carried out optical measurements and analyzed the data, assisted by P.-C.H, and S.-J.C. C.-K.Y., T. C., Y.-W.H, and X.-W. Z performed theoretical analysis, assisted by P.-C.H. X.-W. Z and T. C performed DFT calculations. J.-Y.Z, P.-C.H, and S.-J.C. fabricated the devices. Y.O. and S.A.T. synthesized TMDC crystals. K.W. and T.T. synthesized hBN crystals. C.-K.Y. wrote the manuscript with inputs from all authors.

**Corresponding Author**

* To whom correspondence should be addressed. Email: chawkyong@phys.ntu.edu.tw, tingcao@uw.edu

**Data Availability Statement**

The data that support the findings of this study are available from the corresponding author upon request.

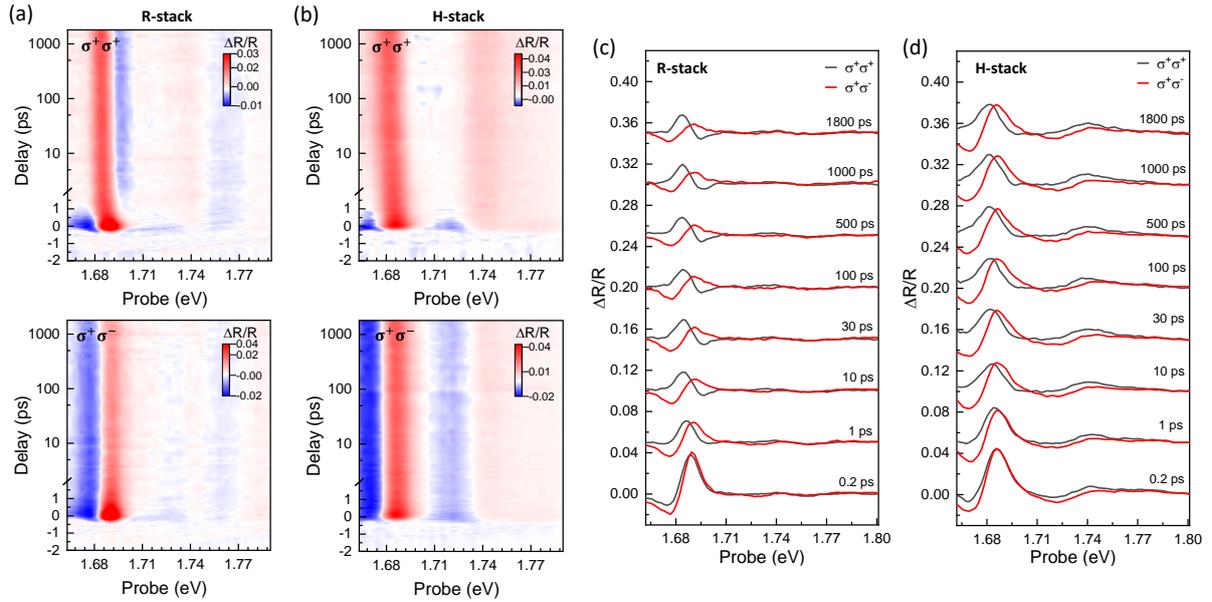

**Extended Data Fig. 1| Ultrafast thermalization of interlayer excitons. a, b.** Two-dimensional plot of transient reflectance spectra of (a) R-stacked and (b) H-stacked $WSe_2/WS_2$ heterobilayer detected using $\sigma^+$ (top panels) and $\sigma^-$ (bottom panels) probe pulses at various delay times after $\sigma^+$ pump excitation. The color scale, vertical axis and horizontal axis represent the relative reflectivity change $\Delta R/R$, the pump-probe time delay $\Delta\tau$, and the probe photon energy, respectively. The positive (negative) $\Delta R/R$ represents a decrease (increase) of absorption. **c, d.** The corresponding line cuts of $\Delta R/R$ spectra at selected $\Delta\tau$ for (c) R-stacked and (d) H-stacked $WSe_2/WS_2$ heterobilayer. The $\Delta R/R$ spectra evolve rapidly within the first few picoseconds, reflecting the ultrafast electron transfer from the $WSe_2$ layer into $WS_2$, and subsequent thermalization of interlayer excitons in the moiré traps. Beyond ∼ 100 ps, the spectral lineshape and intensity remain unchanged, indicating that excitons have fully relaxed to the band edge with negligible recombination loss.

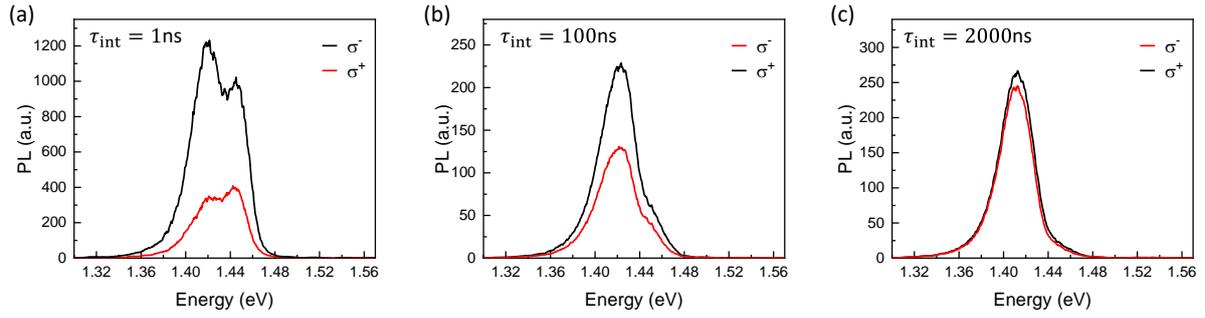

**Extended Data Fig. 2| Helicity-resolved PL spectra as a function of integration time.** Transient PL spectra of H-stacks recorded at a fixed delay $\Delta\tau = 1$ ns, with varying integration windows $\tau_{int}$. (a) For $\tau_{int} = 1$ ns, both the single-occupancy (IX$_1$) and doublon (IX$_2$) emission lines are clearly resolved, and exhibit strong valley polarization. (b) At $\tau_{int} = 100$ ns, IX$_2$ emission is significantly weaker relative to IX$_1$, reflecting the faster decay of doublons. (c) For $\tau_{int} = 2000$ ns, corresponding to integration over the full PL emission following a single excitation pulse, only weakly polarized IX$_1$ emission remains, as both doublon and valley-polarization lifetime are much shorter than the integration time.

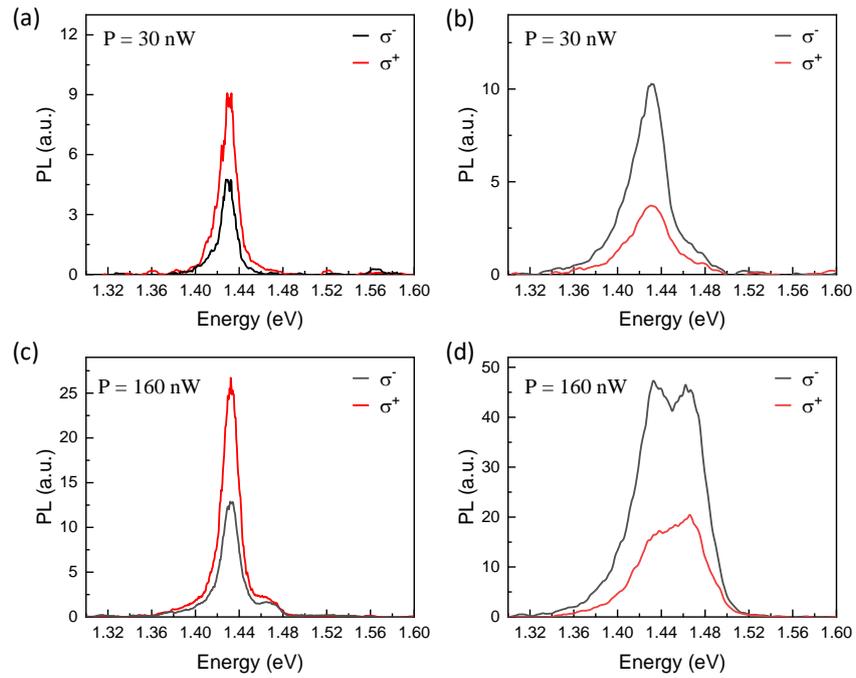

**Extended Data Fig. 3| Helicity-resolved PL spectra on other sets of heterobilayers.** Transient PL spectra recorded at a fixed delay $\Delta\tau = 1$ ns at excitation power of 30 nW (a, b) and 160 nW (c, d) for another set of R-stack and H-stack heterobilayers. In both samples, IX$_2$ emission can be observed when the excitation power is set to 160 nW. This peak shows clear valley-polarization in H-stack but not in R-stack.

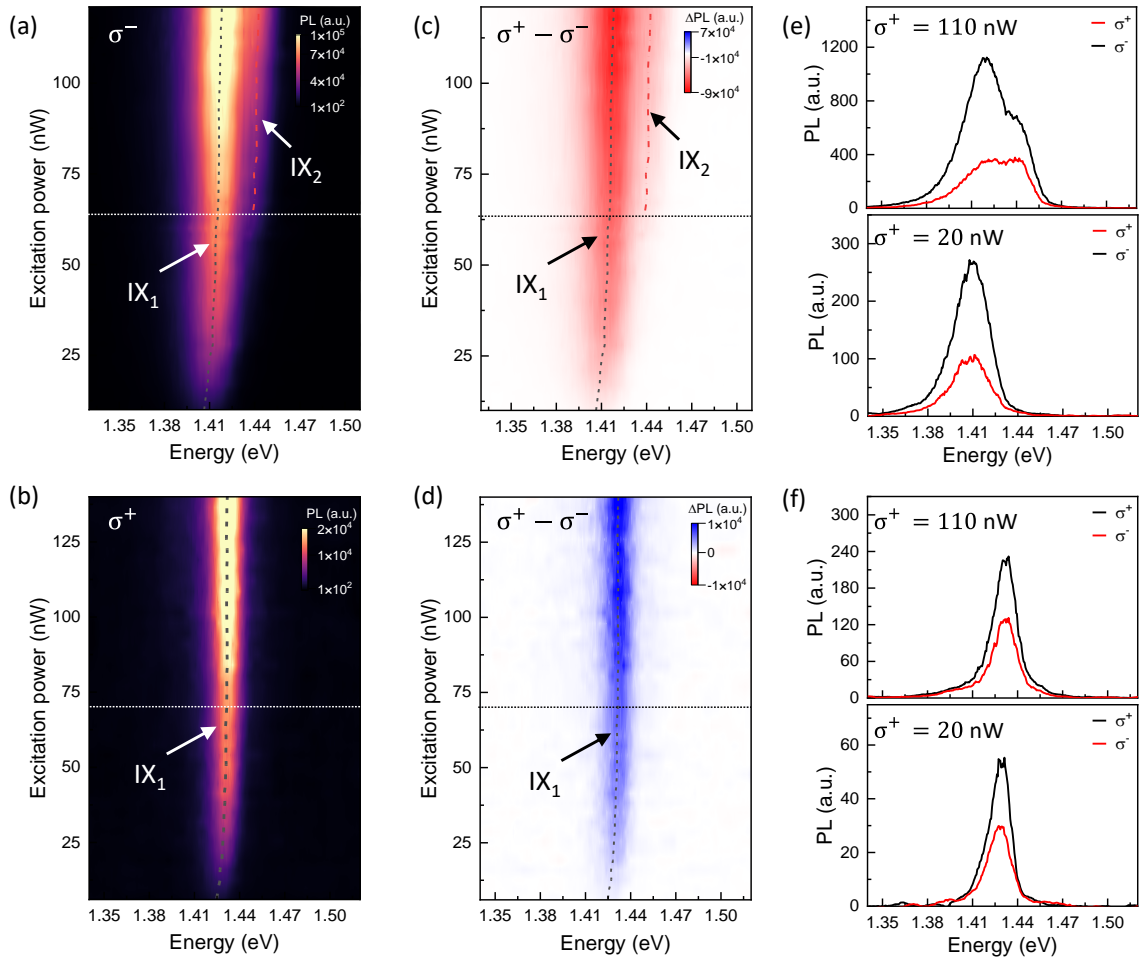

**Extended Data Fig. 4| Helicity-resolved exciton recombination at $\Delta\tau = 6$ ns. a, b.** Transient PL spectra for (a) H-stacked and (b) R-stacked WSe$_2$/WS$_2$ heterobilayers, recorded at $\Delta\tau = 6$ ns after $\sigma^+$-pump excitation. In the H-stack, an additional IX$_2$ emission appears once the excitation power exceeds ~70 nW, corresponding to filling factor $\bar{\nu}_{ex} \approx 1$, in addition to the primary IX$_1$ line. In contrast, the R-stack exhibits only IX$_1$ emission over the entire excitation range, indicating rapid Auger annihilation of doublons. Horizontal dotted lines mark the $\bar{\nu}_{ex} = 1$ threshold, and vertical markers trace the IX$_1$ and IX$_2$ peak positions. **c, d.** Helicity contrast ($\sigma^+ - \sigma^-$) at $\Delta\tau = 6$ ns for (c) H-stack and (d) R-stack. IX$_1$ retains strong valley polarization in both stackings, whereas IX$_2$ shows weak valley polarization only in H-stacked heterobilayers. **e-f.** Helicity-resolved PL spectra at excitation powers of 20 nW and 110 nW for (e) H-stacked- and (f) R-stacked WSe$_2$/WS$_2$, respectively.

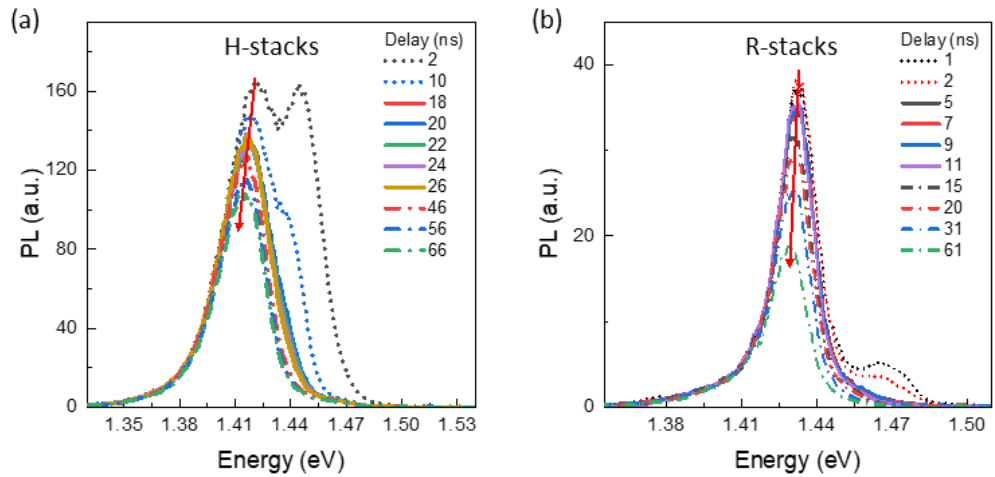

**Extended Data Fig. 5| PL spectra at different delays. a, b.** Transient PL spectra at various delays for (a) H-stack and (b) R-stack. The excitation power is set to 115 nW ($\bar{\nu}_{ex} = 1.5$). The emission spectra with dotted lines show $IX_1$ and $IX_2$ transitions. After $IX_2$ recombination, the emission is dominated by $IX_1$ and the lineshapes (solid line spectra) remain essentially unchanged over prolonged periods, consistent with the emergence of the Mott plateau. At long delay, the emission (dash-dot lines) reduces and redshifts, suggesting the melting of Mott plateau.

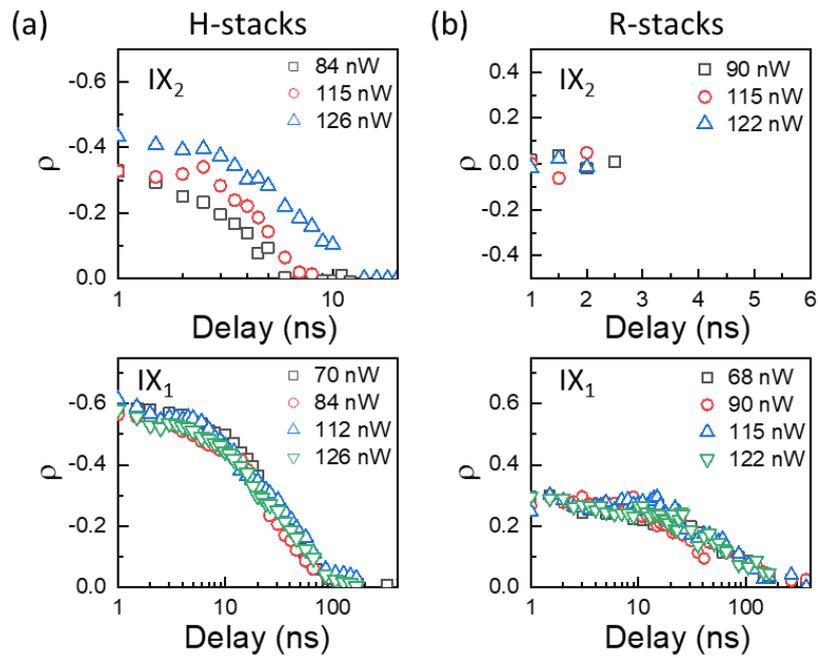

**Extended Data Fig. 6| Degree of circular polarization. a, b.** Degree of circular polarization $\rho$ at various excitation fluences for (a) H-stack and (b) R-stack. The $\rho$ of $IX_1$ persists over tens of nanoseconds and is independent of the excitation density. By contrast, $IX_2$ polarization appears only in H-stack and strengthens with doublon density.